\documentclass[preprint,showpacs,preprintnumbers,amsmath,amssymb]{revtex4}
\usepackage{graphicx}

\begin{document}

\thispagestyle{empty}

\title{Reduction of the Casimir force from indium tin oxide film by UV treatment}

\author{C.-C.~Chang,${}^1$ A.~A.~Banishev,${}^1$
G.~L.~Klimchitskaya,${}^2$
 V.~M.~Mostepanenko,${}^3$
and U.~Mohideen${}^1$}

\affiliation{
${}^1$Department of Physics and Astronomy,
University of California, Riverside, California 92521,
USA\\
${}^2$North-West Technical University, Millionnaya Street 5, St.Petersburg,
191065, Russia \\
${}^3$Noncommercial Partnership
``Scientific Instruments'',  Tverskaya Street 11, Moscow,  103905, Russia
}

\begin{abstract}
A significant decrease in the magnitude of the Casimir force
(from 21\% to 35\%) was observed after an
indium tin oxide (ITO) sample interacting
with an Au sphere was subjected to the UV treatment.
Measurements were performed by using an atomic force
microscope (AFM) in high vacuum. The experimental results are
compared with theory, and a  hypothetical explanation for the
observed phenomenon is proposed.
\end{abstract}

\maketitle

The Casimir effect \cite{1} is important in various
fields from condensed matter physics and
nanotechnology to atomic physics and the theory of fundamental
interactions \cite{2,3,4,5,6,7,8}.
Recent experimental progress \cite{9} has allowed alteration
of the magnitude of the Casimir force by using test bodies
made of materials other than metals. This is of major
importance for the problem of stiction \cite{3} in
microelectromechanical devices,
where it is desirable to make all background forces as small as
possible. Thus, it was demonstrated \cite{10,11} that the Casimir
force between an Au sphere and a Si plate is smaller by
25\%--40\% in comparison with the case of two Au test bodies.
For a plate made of ITO
interacting with an Au sphere, the gradient of the Casimir force
was shown \cite{12} to be roughly 40\%--50\% smaller than between
an Au sphere and an Au plate. For an Au sphere interacting with
AgInSbTe plate  the gradient of the Casimir force decreases in
magnitude by approximately 20\% when the crystalline plate is
replaced with an amorphous plate \cite{13}. The gradient of the
Casimir force between an Au sphere and a semimetallic plate
was measured to be about 25\%--35\% smaller than that for both
Au bodies \cite{14}. In addition, the increase in
magnitude of the Casimir force between an Au sphere and a Si
plate by a few percent was observed when the plate was
illuminated with laser pulses \cite{15}. In all these cases
measured changes in the force were produced by respective
modifications in the optical properties of plates.

In this Letter we report a striking phenomenon of a pronounced
decrease in the magnitude of the Casimir force between an
Au-coated sphere and ITO film deposited on a quartz substrate
after the film undergoes a UV treatment. It is surprising
that the observed decrease is not associated with corresponding
modification in the optical properties of the film sufficient
for changing the Casimir force in accordance with
the Lifshitz theory. The hypothetical explanation of this
phenomenon, which could find multidisciplinary applications,
is provided.
%%%%%%%%%%%%%%%%%%%%%%%%%%%%%%%%%%%%%%%%%%%%%

Measurements of the Casimir force were performed using
a modified  multimode AFM.
It was  placed in a high vacuum chamber. Only oil free mechanical
and turbo pumps were used to obtain the vacuum. The experiments were
done at a pressure of $10^{-6}\,$Torr. The AFM was modified to be
free of volatile organics. To have a low vibration noise environment,
we used an optical table and a sand damper box to prevent coupling
of the low frequency
noise from the mechanical and turbo pumps. To stabilize the
AFM laser used for the detection of cantilever deflection, we employed
a liquid nitrogen cooling system, which maintained the temperature
of the AFM at 2${}^{\circ}$C. The cooling system reduced the laser
noise and drift. It also served as an additional cryo pump to obtain
the high vacuum.

A polystyrene sphere  was glued with silver epoxy
($20\,\mu\mbox{m}\times 20\,\mu\mbox{m}$ spot) to the tip of a triangle
silicon nitride cantilever with a nominal spring constant
$\sim 0.01\,$N/m. The cantilever-sphere system was then
coated with a 10\,nm Cr layer, followed by 20\,nm Al layer and finally
with a $105\pm 1\,$nm Au layer in an oil free thermal evaporator
with a $10^{-7}\,$Torr vacuum. To make sure that the Au surface is
smooth, the coatings were done at a very low deposition rate of
3.75\,{\AA}/min.  The radius of Au sphere was determined using a
SEM to be $101.23\pm 0.5\,\mu$m after the end of the force measurements.

The ITO films  were prepared by RF sputtering (Thinfilms Inc.)
on a 1 cm square  single crystal quartz plates of 1\,mm
thickness.
A film thickness was measured to be  $74.6\pm 0.2\,$nm.
The nominal film resistivity is $42\,\Omega/$sq.
The ITO film on the quartz plate was cleaned with the following
procedure: First, the plate was immersed in acetone and cleaned in
an ultrasonic bath for 15\,min. It was then rinsed 3 times in DI water.
 This ultrasonic cleaning procedure and water rinsing was repeated next
with methanol followed by ethanol. After ethanol cleaning the sample
was dried in a flow of pure nitrogen gas. Next, electrical contacts
to copper wires were made by soldering with indium wire.
Then the force measurements described below were
performed. Thereafter the same sample was
UV treated, and the measurements were repeated.

To prepare the UV-treated sample, the ITO film on a quartz plate
was placed in a special air chamber containing a UV lamp. A pen-ray
Mercury lamp with a length of 9.0${}^{\prime\prime}$ and outside
diameter of 0.375${}^{\prime\prime}$ was used as the UV source.
The lamp emits a spectrum with the primary peak at 254\,nm
(5.4\,mW/cm${}^2$ at 1.9\,cm distance) and a secondary peak at
365\,nm (0.2\,mW/cm${}^2$ at 1.9\,cm distance).
The sample was placed 1\,cm from the lamp for 12 hours. The
UV-treated sample was cleaned as above. The roughness profiles
of the Au coating on the sphere and the ITO film were measured
with an AFM (the variances are 3.17 and 2.28\,nm,
respectively).

The measurement procedure is as follows.
The samples were inserted on top of the AFM piezo and placed
in the high vacuum chamber.
The ITO film was connected to voltage supply (33120A, Agilent Inc.)
operating with 1\,$\mu$V resolution. A 1\,k$\Omega$  resistor was
connected in the series with the voltage supply to prevent
surge currents and protect the sample surface during sphere-plate
contact. The cantilever-sphere system was mounted on the AFM
head which was connected to the ground.
To reduce the electrical noise, care was taken to make Ohmic
contacts and eliminate all Schottky barriers, to the ITO plate
and Au sphere.  To minimize electrical ground loops all the
electrical ground connections were unified to the AFM ground.
Ten different voltages $V_i$ ($i=1,\,2,\,\ldots,\,10$)
in the range from --260 to --100\,mV
(from --25 to 150\,mV for the UV-treated sample) were applied to
the plate, while the sphere remained grounded.

The total force $F_{\rm tot}$ (electrostatic plus Casimir)
between the Au sphere and the ITO plate was measured as a
function of sphere-plate separation. A 0.05\,Hz continuous
triangular voltage was applied to the AFM piezo to change
the separations between the sphere and the plate by 2\,$\mu$m.
The interferometric calibration of the piezo was done as
discussed previously \cite{i1,i2}.  Starting at the maximum
separation, the plate was moved towards the sphere and the
corresponding cantilever deflection was recorded every
0.2\,nm till the plate contacted the sphere.
The sphere-plate separation $a$ is given by:
\begin{equation}
a=z_{\rm piezo}+S_{\rm def}m+z_0,
\label{eqI1}
\end{equation}
\noindent
where $z_{\rm piezo}$ is the movement of the plate due
to the piezo, the product of deflection signal $S_{\rm def}$
and deflection coefficient $m$
is the change in separation distance due to
 cantilever deflection, and $z_0$
is the average separation on contact due to surface roughness.

After the contact of the sphere and the plate, the
cantilever-sphere system vibrates with a large amplitude.
To allow time for this vibration to damp out, a 5\,s delay
was introduced after every cycle of data acquisition.
The total force measurement was repeated 10 times for every
applied voltage.

The electrostatic calibration was performed as follows.
The total force is given by
$F_{\rm tot}=kmS_{\rm def}$,  where $k$ is cantilever spring
constant.
Any linear deflection with sphere-plate separation due to
mechanical drift of the photodetector was first subtracted.
For separations larger than 1.7\,$\mu$m, the total
force between the sphere and the plate is below the
instrumental sensitivity. The small observed deflection
from the region 1.7--2\,$\mu$m was fitted to a straight
line and the coefficients were used to subtract this drift
from the whole force curve for all separations.
The sphere-plate contact was corrected for mechanical drift
and the cantilever deflection coefficient
$m=104.4\pm 0.5\,$nm per unit deflection signal and
$m=103.5\pm 0.6\,$nm per unit deflection signal
 was determined as discussed in Ref.~\cite{i3} for
the untreated and UV-treated samples, respectively.
These values of $m$ were used to
determine the sphere-plate separation $a$ in Eq.~(\ref{eqI1})
up to the value of $z_0$ (which is a constant for the complete
set of measurements).

The residual potential difference $V_0$ between the Au sphere
and ITO plate, the spring constant $k$ and the average separation
on contact $z_0$ were determined from the parabolic dependence
of the total force on the applied voltage.
The cantilever deflection at every applied voltage was determined
at intervals of 1\,nm sphere-plate separations using linear
interpolation.  For each separation the $S_{\rm def}$  was
plotted as a function of the applied voltage.
{}From the parabolas generated, the residual potential $V_0$,
which corresponds to the value of the voltage for zero electrostatic
force at the vertex of the parabola, was determined by a least
$\chi^2$ fitting procedure.
The values $V_0= -196.8\pm 1.5\,$mV for untreated and
$V_0=65\pm 2\,$mV for UV-treated sample were found to be
independent of separation.
The curvature of the parabola was also determined at every
separation $a$. This curvature as a function of the sphere-plate
separation was fitted to the spatial-dependent part of the
electrostatic force to obtain the average separation on
contact and the spring constant. As a result,
$z_0= 29.5\pm 0.4\,$nm,  $k=0.0139\pm 0.0001\,$N/m for the untreated
sample, and
$z_0=29\pm 0.6\,$nm, $k=0.0138\pm 0.0001\,$N/m  for the UV-treated sample.
These values of $z_0$ were used to determine absolute separations
from Eq.~(\ref{eqI1}). The values of $k$ and $m$ were used
to convert the measured $S_{\rm def}$  to values of force.

%%%%%%%%%%%%%%%%%%%%%%%%%%%%%%%%%%%%%%%%%%%%%%%
Using the above procedure, we have measured the total force
at each of the applied voltages
$V_i$  over the separation region
from 60 to 300\,nm with a step of 1\,nm.
This measurement was repeated ten times resulting in one
hundred values of the total force $F_{ik}^{\,\rm tot}(a)$
($k=1,\,2,\,\ldots,\,10$) at each separation.

The systematic error in the values of total force measured,
$\Delta_sF_{ik}^{\,\rm tot}(a)$, is added from the
separation-dependent calibration error and the instrumental noise
including background noise level which does not depend on
separation. The resulting $\Delta_sF_{ik}^{\,\rm tot}(a)$
is equal to 2.1, 1.5, and 1.1\,pN at
 $a=60$, 100 and $\geq 200\,$nm, respectively
(all errors here and below are determined at a 95\% confidence
level).

The values of the Casimir force at each separation were obtained
by the subtraction of the electric force in the sphere-plate
geometry \cite{2}
\begin{equation}
F_{ik}(a)=F_{ik}^{\,\rm tot}(a)-2\pi\epsilon_0(V-V_0)^2
\sum_{n=1}^{\infty}\frac{\coth\alpha-n\coth{n\alpha}}{\sinh{n\alpha}},
\label{eq1}
\end{equation}
\noindent
where $\cosh\alpha=1+a/R$, and $\epsilon_0$ is the permittivity
of
vacuum. The systematic errors in the electric force are
determined by the errors in $V_0$, $R$ and, primarily,
in $a$. The resulting systematic error of the Casimir force,
was found from the combination of errors in the
total and electric forces.
The variance of the mean Casimir force from 100 repetitions
does not depend on $a$ and is equal to 0.55\,pN.
Then from the Student distribution the random error was
obtained, which was combined with the systematic error to
find the total experimental error, $\Delta^{\!\rm tot}F(a)$,
in the measured Casimir force.

In Fig.~1, the mean Casimir force measured as a function of
separation is shown as crosses for an untreated
 and
UV-treated ITO films (lower and upper sets of crosses,
respectively). The arms of the crosses indicate
the total experimental errors in the measurement of separations
and forces.
To illustrate, in the inset to Fig.~2 we plot
$\Delta^{\!\rm tot}F(a)$
 for the UV-treated sample as a function of separation.
The lower set of crosses in Fig.~1 indicates a
 40\%--50\% decrease
in the force magnitude as compared to the case
of two Au bodies (e.g., at $a=80\,$nm the measured force is
--144\,pN in comparison with --269\,pN), in agreement
with Ref.~\cite{12}.

{}From Fig.~1 it can be seen that the magnitude of the Casimir
force from the UV-treated film is markedly less than
from the untreated one.
The relative decrease in the force magnitude is equal to 21\%
at $a=60\,$nm, increases to approximately 35\% at $a=130\,$nm,
and preserves this value at larger separations.
Measurements were repeated several times with different
samples, untreated and UV-treated, leading to similar
results.

The measurement results were compared with computations using
the Lifshitz theory at the experimental temperature
$2^{\circ}$C for an ITO film on
a quartz substrate interacting with an Au sphere, and the
proximity force approximation (PFA). The error in using the
PFA for real materials with given experimental parameters
does not exceed 0.3\% \cite{T16,T17,T18}. The roughness
of both ITO and Au surfaces
was  taken into account by means
of geometrical averaging \cite{2,9}.
The resulting correction to the force is equal to 2.2\%
at $a=80\,$nm, and becomes less than 1\% and 0.5\%
at $a\geq 90\,$nm and $a\geq 116\,$nm, respectively.
The dielectric properties of ITO films were investigated
using the untreated and UV-treated samples prepared in the same
way and under the same conditions as those used in
measurements of the Casimir force. The imaginary part of
$\varepsilon_{\rm ITO}(\omega)$ was determined \cite{T19} in
the frequency region from 0.04\,eV to 8.27\,eV with IR-VASE
and VUV-VASE ellipsometers at low and high frequencies,
respectively. The results for ${\rm Im}\,\varepsilon$
are shown in the insets to Fig.~3(a,b) by the solid lines
for untreated and UV-treated samples, respectively.
They were extrapolated to lower frequencies by the Drude
model with the plasma frequency $\omega_p=1.5\,$eV and
relaxation parameters $\gamma=0.128\,$eV and $\gamma=0.132\,$eV.
 The dashed lines in
the insets show the limits of a possible smooth extrapolation
of the measured optical data to higher frequencies obtained
from the oscillator model.
As can be seen in the insets, ${\rm Im}\,\varepsilon_{\rm ITO}$
is only slightly affected by the UV treatment.
The respective dielectric
permittivities along the imaginary frequency axis,
$\varepsilon_{\rm ITO}(i\xi)$, are shown in main part of
Fig.~3(a,b) by solid and dashed lines. Here, the
dashed [in Fig.~3(a)] and solid [in Fig.~3(b)] lines indicate
$\varepsilon_{\rm ITO}(i\xi)$ with the role of charge
carriers neglected.
For quartz, the $\varepsilon(i\xi)$ of Ref.~\cite{T20} was used.
The optical properties of Au were taken from Ref.~\cite{T21}
and extrapolated to lower frequencies by means of the Drude
model with $\omega_p=9.0\,$eV and $\gamma=0.035\,$eV
(this extrapolation nicely fits \cite{T22} the
measured data \cite{T21}).

The theoretical results are shown by the lower and upper bands
between the pairs of solid lines in the inset to Fig.~1
for the untreated and UV-treated samples, respectively.
In computations, we have used the pairs of solid lines in both
Figs.~3(a) and (b), i.e., neglected the contribution of
charge carriers in the UV-treated film. In the main field of
Fig.~2, we repeat the experimental data for the UV-treated
sample, but show the theoretical results computed using the dashed
lines in Fig.~3(b), i.e., with account of charge carriers.
As can be seen in Figs.~1 and 2, the data are in excellent
agreement with theory with charge carriers of the UV-treated
sample neglected, but at a 95\% confidence level
exclude the theory taking these charge carriers
into account. Note that the same results are obtained when Au is
described by the generalized plasma-like model \cite{2,9}.
One can hypothesize that the UV-treatment resulted in the
transition of an ITO film to a dielectric state without
noticeable change of its optical properties at room temperature
(according to Ref.~\cite{T23}, the UV-treatment of ITO leads
to lower mobility of charge carriers). This hypothesis could be
verified in future by the investigation of electrical properties
at very low temperature. Then, the neglect of charge carriers
for a UV-treated sample fits  commonly accepted practice in the
application of the Lifshitz theory to dielectric
bodies \cite{2,9}.

In the foregoing, we have experimentally demonstrated that the
UV-treatment of an ITO sample interacting with an Au sphere
leads to an overall decrease of the Casimir force up to 65\%
in comparison to Au-Au test bodies. This result is of much
practical importance for  addressing problems of lubrication and
stiction in microelectromechanical systems.
The hypothetical explanation of the observed
phenomenon provided invites further investigation of the
physical properties of complicated chemical compounds
including their interaction with
zero-point and thermal fluctuations of the electromagnetic
field.

This work was supported by the DARPA Grant under Contract
No.~S-000354 (equipment, A.B., U.M.),  NSF Grant
No.~PHY0970161 (C.-C.C, G.L.K., V.M.M., U.M.) and DOE Grant
No.~DEF010204ER46131 (G.L.K., V.M.M., U.M.).

%%%%%%%%%%%%%%%%%%%%%%%%%%%%%%%%%%

%%%%%%%%%%%%
%%%%%%%%%%%%
%\end{document}
\newpage
%%%%%%%%%%%%%%%%%%%%%%%%%%%%%%%%%%%%%%%%%%%%
%%%___FIGURES___%%%%%%%%%%%%%%%%%%%%%%%%%%%%
%%%%%%%%%%%%%%%%%%%%%%%%%%%%%%%%%%%%%%%%%%%%
%%%%_____FIG__1___%%%%%%%%%%%%%%%%%%%%%%%%%
\begin{figure*}[h]
\vspace*{-12.cm}
\centerline{\hspace*{2cm}
\includegraphics{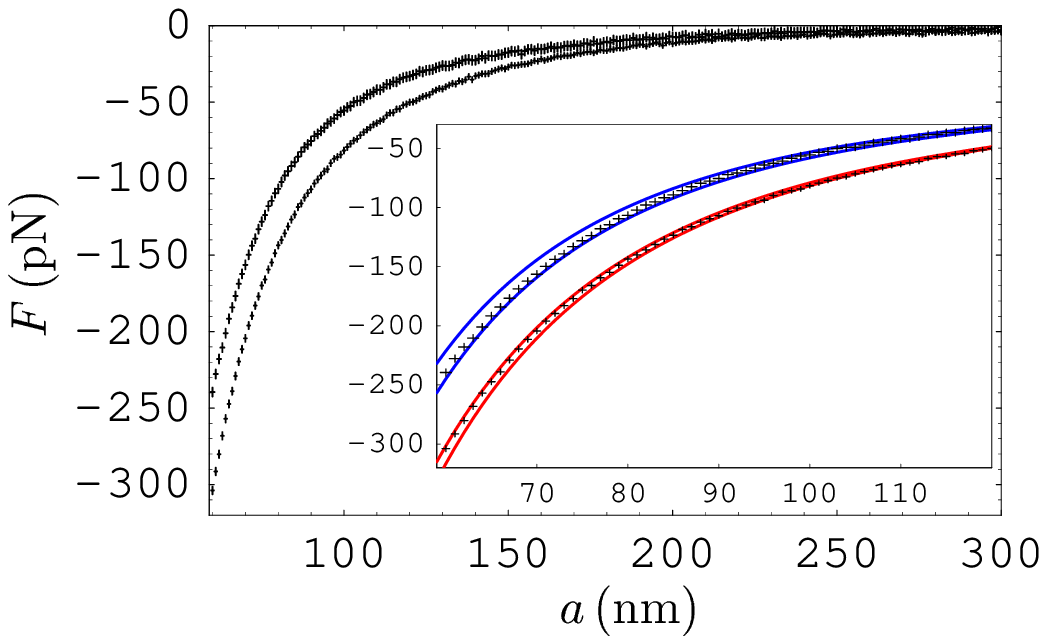}
}
\vspace*{-6.5cm}
\caption{The mean measured Casimir forces between an Au sphere
and untreated and UV-treated ITO samples as functions of
separation are shown by the lower and upper sets of crosses,
respectively. In the inset the same data are compared with
theory (see text for further discussion).
}
\end{figure*}
%%%%%%%%%%%%%%%%%%%%%%%%%%%%%%%%%%%%%%%%%%%%%%%%%%%%%%%%%
%%%%_____FIG__2___%%%%%%%%%%%%%%%%%%%%%%%%%
\begin{figure*}[h]
\vspace*{-4.cm}
\centerline{\hspace*{2cm}
\includegraphics{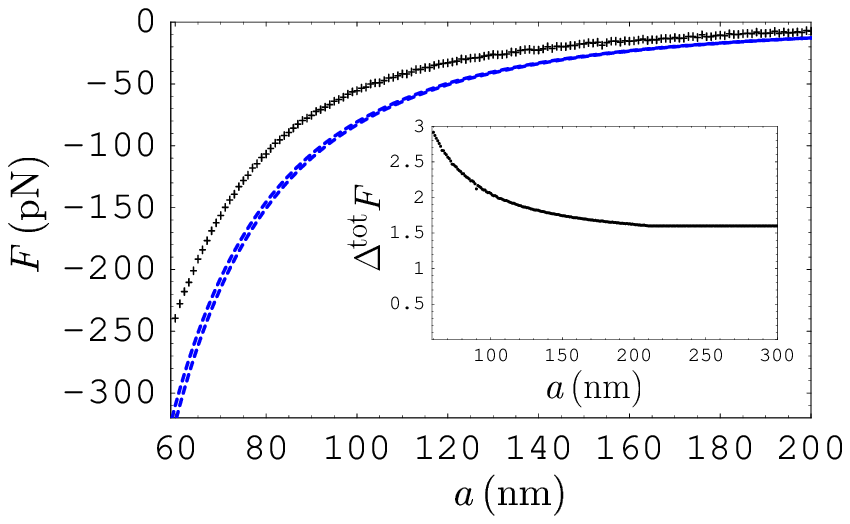}
}
\vspace*{-15.5cm}
\caption{The measured Casimir force between an Au sphere
and an UV-treated ITO sample as functions of
separation is shown as crosses. The band between the dashed
lines indicates the theoretical prediction where the ITO charge
carriers are described by the Drude model.
In the inset, the dependence of the total experimental error
on separation is plotted.
}
\end{figure*}
%%%%%%%%%%%%%%%%%%%%%%%%%%%%%%%%%%%%%%%%%%%%%%%%%%%%%%%%%
%%%%_____FIG__3___%%%%%%%%%%%%%%%%%%%%%%%%%
\begin{figure*}[h]
\vspace*{-2.cm}
\centerline{\hspace*{2cm}
\includegraphics{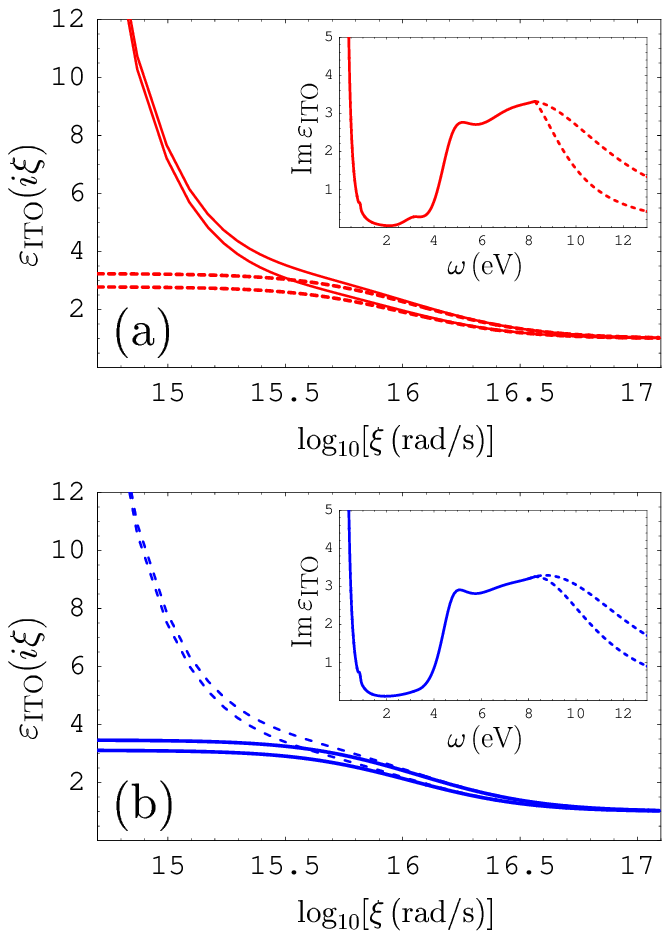}
}
\vspace*{-16.5cm}
\caption{The dielectric permittivities as a function of imaginary
frequency are shown (a) for a untreated ITO sample
with charge carriers included (solid lines) and neglected
(dashed lines) and (b) for a UV-treated sample
with charge carriers included (dashed lines) and neglected
(solid lines). In the insets the imaginary parts of
corresponding dielectric permittivities obtained from
ellipsometry are presented.
}
\end{figure*}
%%%%%%%%%%%%%%%%%%%%%%%%%%%%%%%%%%%%%%%%%%%%%%%%%%%%%%%%%

\begin{thebibliography}{99}
\bibitem {1}
H.~B.~G.~Casimir,
{ Proc. K. Ned. Akad. Wet. B}
{\bf 51}, 793 (1948).
\bibitem{2}
M.~Bordag, G.~L.~Klimchitskaya, U.\ Mohideen, and
V.\ M.\ Mostepanenko, {\it Advances in the Casimir Effect}
(Oxford University Press, Oxford, 2009).
\bibitem{3}
E.~Buks and M.~L.~Roukes,
Phys. Rev. B {\bf 63}, 033402 (2001).
\bibitem{4}
A.~W.~Rodriguez, F.~Capasso, and S.~G.~Johnson,
Nature Photon. {\bf 5}, 211 (2011).
\bibitem{5}
A.~Ashourvan, M.~Miri, and
R.\ Golestanian,
Phys. Rev. Lett. {\bf 98}, 140801 (2007).
\bibitem{6}
H.~B.~Chan et al.,
Phys. Rev. Lett. {\bf 101}, 030401 (2008).
\bibitem{7}
J.~F.~Babb,
Adv. At. Mol. Opt. Phys. {\bf 59}, 1 (2010).
\bibitem{8}
E.~G.~Adelberger, B.~R.~Heckel, and A.~E.~ Nelson,
 Ann. Rev. Nucl. Part. Sci. {\bf 53}, 77 (2003).
\bibitem{9}
G.~L.~Klimchitskaya, U. Mohideen, and V.\ M.\ Mostepanenko,
 Rev. Mod. Phys. {\bf 81}, 1827 (2009);
 Int. J. Mod. Phys. B {\bf 25}, 171 (2011).
\bibitem{10}
F.~Chen et al.,
Phys. Rev. A  {\bf 72}, 020101(R) (2005); {\bf 74}, 022103 (2006).
\bibitem{11}
F.~Chen et al.,
Phys. Rev. Lett. {\bf 97}, 170402 (2006).
\bibitem{12}
S.~de~Man et al.,
Phys. Rev. Lett. {\bf 103}, 040402 (2009);
S.~de~Man, K.~Heeck and D.~Iannuzzi,
Phys. Rev. A {\bf 82}, 062512 (2010).
\bibitem{13}
G.~Torricelli et al.,
Phys. Rev. A {\bf 82}, 010101(R) (2010).
\bibitem{14}
G.~Torricelli et al.,
Europhys. Lett. {\bf 93}, 51001 (2011).
\bibitem{15}
F.~Chen et al.,
Optics Express  {\bf 15}, 4823 (2007);
Phys. Rev. B  {\bf 76}, 035338 (2007).
%%%%%%%%%%%%%%%%%%%%%%%%%%%%%%%%%
\bibitem{i1}
F.~Chen and U.~Mohideen,
Rev. Sci. Instrum. {\bf 72}, 3100 (2001).
\bibitem{i2}
H.~E.~Grecco and O.~E.~Martinez,
Appl. Opt. {\bf 41}, 6646 (2002).
\bibitem{i3}
H.-C.\ Chiu et al.,
J. Phys. A {\bf 41}, 164022 (2008).
%%%%%%%%%%%%%%%%%%%%%%%%%%%%%%%%%
\bibitem{T16}
A.~Canaguier-Durand et al.,
Phys. Rev. Lett. {\bf 104}, 040403 (2010).
\bibitem{T17}
R.~Zandi, T.~Emig, and U.~Mohideen,
Phys. Rev. B {\bf 81}, 195423 (2010).
\bibitem{T18}
B.~Geyer, G.~L.~Klimchitskaya, and V.~M.~Mostepanenko,
Phys. Rev. A  {\bf 82}, 032513 (2010).
\bibitem{T19}
http://www.jawoollam.com
\bibitem{T20}
L.~Bergstr\"{o}m,
Adv. Coll. Interface Sci. {\bf 70}, 125 (1997).
\bibitem {T21}
{\it Handbook of Optical Constants of Solids},
ed. E.~D.~Palik (Academic, New York, 1985).
\bibitem {T22}
G.~Bimonte,
Phys. Rev. A {\bf 83}, 042109 (2011).
\bibitem {T23}
C.~N.~Li et al.,
Appl. Phys. A {\bf 80}, 301 (2005).
\end{thebibliography}
\end{document}